\documentclass{article}
\usepackage{dafne99pro}
\usepackage{epsf}
\newcommand{\bea}{$$\begin{array}{rcl}}
\newcommand{\eea}{\end{array}$$}

\def\vev#1{\left\langle #1\right\rangle}
\def\Im{\mathop{\mbox{Im}}}

\def\ee{$\varepsilon'/\varepsilon$~}
\begin{document}
\title{ 
THE $\Delta I = 1/2$ RULE AND $\varepsilon'/\varepsilon$ IN THE
CHIRAL~QUARK~MODEL
}
\author{
Stefano Bertolini        \\
{\em INFN, Sezione di Trieste and SISSA} \\ 
{\em via Beirut 4, I-34013 Trieste, Italy}
}
\maketitle
\baselineskip=11.6pt
\begin{abstract}
I discuss the role of the $\Delta I = 1/2$ selection rule in
$K\to\pi\pi$ decays for the theoretical calculations of \ee. 
Lacking reliable ``first principle'' calculations, 
phenomenological approaches may help in understanding 
correlations among different contributions and available experimental data.
In particular, in the chiral quark model approach the same dynamics which
underlies the $\Delta I = 1/2$ selection rule in kaon decays 
appears to enhance the $K\to\pi\pi$ matrix elements of
the gluonic penguins, thus driving \ee in the range of the
recent experimental measurements.
\end{abstract}
\baselineskip=14pt
\vskip 0.5 in

The results announced this year by the KTeV\cite{KTeV} 
and NA48\cite{NA48} collaborations
have marked a great experimental achievement,
establishing 35 years after the discovery of CP violation
in the neutral kaon system\cite{Christenson}
the existence of a much smaller violation acting directly in the
decays. 

While the Standard Model (SM) of strong and electroweak interactions
provides an economical and elegant understanding
of indirect ($\varepsilon$) and direct ($\varepsilon'$) 
CP violation in term of a single phase,
the detailed calculation of the size of these effects 
implies mastering strong interactions at a scale 
where perturbative methods break down. In addition, 
CP violation in $K\to\pi\pi$ decays
is the result of a destructive interference between
two sets of contributions,
which may inflate up to an order of magnitude the uncertainties
on the individual hadronic matrix elements of the effective 
four-quark operators.
THis makes predicting \ee a complex and subtle theoretical
challenge\cite{review}.

\begin{figure}[t]
\epsfxsize=9cm
\centerline{\epsfbox{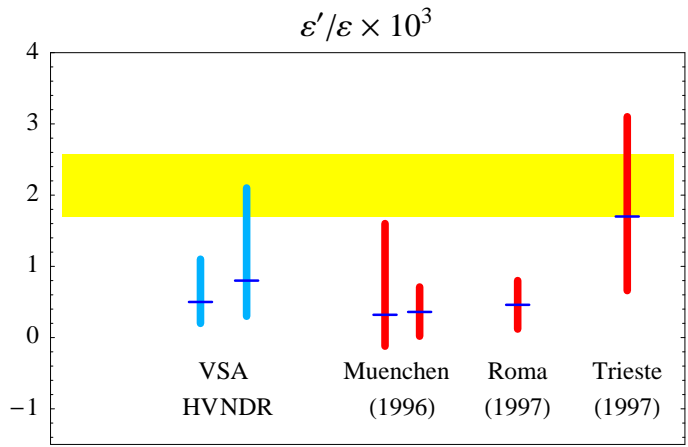}}
\caption{\it 
The combined 1-$\sigma$ average of the
NA31, E731, KTeV and NA48 results (\ee = $21.2\pm 4.6\times 10^{-4}$)
is shown by the gray horizontal band (the error is inflated
according to the Particle Data Group procedure when averaging over
data with substantially different central values).  
The old M\"unchen, Roma and Trieste theoretical predictions for \ee are
depicted by the vertical bars with their central values.
For comparison, the VSA estimate is shown using two renormalization schemes.}
\label{fig3new}
\end{figure}

In Fig.~\ref{fig3new} I summarize the comparison 
of the theoretical 
predictions available before 
the KTeV announcement early this year with the present
experimental data.
The gray horizontal band shows
the one-sigma average of the old NA31 (CERN) and E731 (Fermilab) data
and the new KTeV and NA48 results. 
The vertical lines show the ranges of the
published theoretical $predictions$ (before February 1999), 
identified with the cities where most members of the groups reside.
The range of the naive Vacuum Saturation Approximation (VSA) is
shown for comparison. 

By considering the complexity of the problem,
the theoretical calculations reported in Fig.~\ref{fig3new}, 
show a remarkable agreement,
all of them pointing to a non-vanishing positive effect in the SM.
On the other hand, if we focus our attention on
the central values,  the M\"unchen (phenomenological $1/N$)
and Rome (lattice) calculations definitely prefer
the $10^{-4}$ regime, contrary to the Trieste result which is 
above $10^{-3}$.

Without entering the details of the calculations, it is important
to emphasize that the abovementioned difference 
is mainly due to the different size of the 
hadronic matrix element of the gluonic penguin $Q_6$
obtained in the various approaches. 
While the M\"unchen and Rome calculations
assume for $\vev{Q_6}$ values in the neighboroud of the 
leading $1/N$ result (naive factorization), the Trieste
calculation, based on 
the effective Chiral Quark Model ($\chi$QM)\cite{Weinberg-GeorgiManohar}
and chiral expansion, finds a substantial enhancement of the
$I=0$ $K\to\pi\pi$ amplitudes, which affect $both$ current-current and
penguin operators. 
The bulk of such an enhancement
can be simply understood in terms of chiral dynamics (final-state interactions)
relating the \ee prediction 
to the phenomenological embedding of the $\Delta I= 1/2$ selection rule.

The $\Delta I = 1/2$ selection rule in $K\to\pi\pi$ decays is known by 
some 45 years\cite{Pais-Gell-Mann} and it states the experimental
evidence that kaons
are 400 times more likely to decay in the $I=0$ two-pion state
than in the $I=2$ component. This rule is not justified by any
general symmetry consideration and, although it is common understanding 
that its explanation must be rooted in the dynamics of strong interactions,
there is no up to date derivation of this effect from first principle QCD. 

As summarized by Martinelli at this conference\cite{Martinelli}
lattice cannot provide us at present with reliable calculations
of the $I=0$ penguin operators relevant to \ee, as well as of the 
$I=0$ components of the hadronic matrix elements of the tree-level
current-current operators (penguin contractions), which are relevant for the
$\Delta I = 1/2$ selection rule.

In the M\"unich approach\cite{Buras} the $\Delta I = 1/2$ rule
is used in order to determine phenomenologically the  
matrix elements of $Q_{1,2}$ and, 
via operatorial relations, some of the matrix elements of the
left-handed penguins. Unfortunately, the approach does not allow
for a phenomenological determination of the matrix elements of the penguin
operators which are most relevant for \ee, namely the gluonic penguin $Q_6$
and the electroweak penguin $Q_8$. 

In the $\chi$QM approach, the hadronic matrix elements
can be computed as an expansion in the external momenta
in terms of three parameters:
the constituent quark mass, the quark condensate
and the gluon condensate.
The Trieste group has computed the $K\to\pi\pi$ matrix elements
of the $\Delta S =1,2$ effective lagrangian up to $O(p^4)$ in the
chiral expansion\cite{ts98a,ts98b}. 

Hadronic matrix elements and short distance Wilson coefficients are 
then matched
at a scale of $0.8$ GeV as a reasonable
compromise between the ranges of validity 
of perturbation theory and chiral lagrangian.
By requiring the $\Delta I = 1/2$ rule to be reproduced
within a 20\% uncertainty one obtains a phenomenological
determination of the three basic parameters of the model.
This step is crucial
in order to make the model predictive, since there is no a-priori argument
for the consistency of the matching procedure. As a matter of
fact, all computed observables turn
out to be very weakly scale (and renormalization scheme)
dependent in a few hundred MeV range around the 
matching scale.   

\begin{figure}
\epsfxsize=9cm
\centerline{\epsfbox{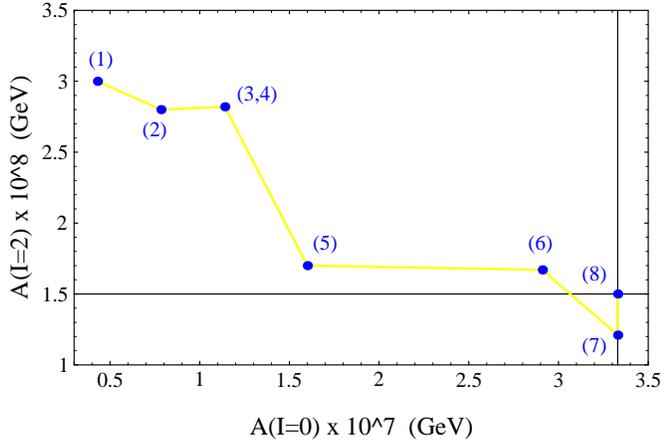}}
\caption{Anatomy of the $\Delta I = 1/2$ rule in the 
$\chi$QM\cite{ts98a}. See the text for explanations.
The cross-hairs indicate the experimental point.}
\label{road}
\end{figure}

Fig.~\ref{road} shows an anatomy of the various contributions
which finally lead to the experimental value
of the $\Delta I = 1/2$ selection rule.

Point (1) represents the result obtained by neglecting QCD
and taking the factorized matrix element for the
tree-level operator $Q_2$, which is the leading electroweak 
contribution. 
The ratio $A_0/A_2$ is found equal
to $\sqrt{2}$: by far off the experimental point (8).
Step (2) includes the effects of perturbative QCD renormalization
on the operators $Q_{1,2}$\cite{Gaillard-etc}. 
Step (3) shows the effect of including the gluonic 
penguin operators\cite{VSZ-GWise-CFGeorgi}. 
Electroweak penguins\cite{Lusignoli-etc} are numerically
negligeable in the CP conserving amplitudes and 
are responsible for the very small shift in the $A_2$ direction.
Perturbative QCD and factorization lead us from (1) to (4).

Non-factorizable gluon-condensate corrections,
a crucial model dependent effect entering at the leading order
in the chiral expansion, produce a substantial
reduction of the $A_2$ amplitude (5), 
as it was first observed by Pich and de Rafael\cite{Pich-deRafael}. 
Moving the analysis to $O(p^4)$
the chiral loop corrections, computed on the LO chiral
lagrangian via dimensional regularization and minimal subtraction, 
lead us from (5) to (6), while 
the finite parts of the NLO counterterms
calculated via the $\chi$QM approach lead us to the point (7).
Finally, step (8) represents the inclusion of $\pi$-$\eta$-$\eta'$ 
isospin breaking effects\cite{Ometapeta}. 

This model dependent anatomy 
shows the relevance of non-factorizable contributions
and higher-order chiral corrections. The suggestion that
chiral dynamics may be relevant to
the understanding of the $\Delta I = 1/2$ selection rule goes back to the
work of Bardeen, Buras and Gerard\cite{BBG} in the $1/N$
framework using a cutoff regularization. This approach
has been recently revived and improved 
by the Dortmund group, with a particular attention
to the matching procedure\cite{Hambye}. 
A pattern similar to that shown
in Fig.~\ref{road} for the chiral loop corrections to $A_0$ and $A_2$
was previously obtained in a NLO chiral
lagrangian analysis, using dimensional regularization, by 
Missimer, Kambor and Wyler\cite{MKWyler}.

The $\chi$QM approach allows us to further investigate the relevance
of chiral corrections for each of the effective quark operators
of the $\Delta S = 1$ lagrangian.
The NLO contributions
to the electroweak penguin matrix elements have been thouroughly studied
for the first time by the Trieste group\cite{ts96a,ts98b}.

\begin{figure}
\epsfxsize=9cm
\centerline{\epsfbox{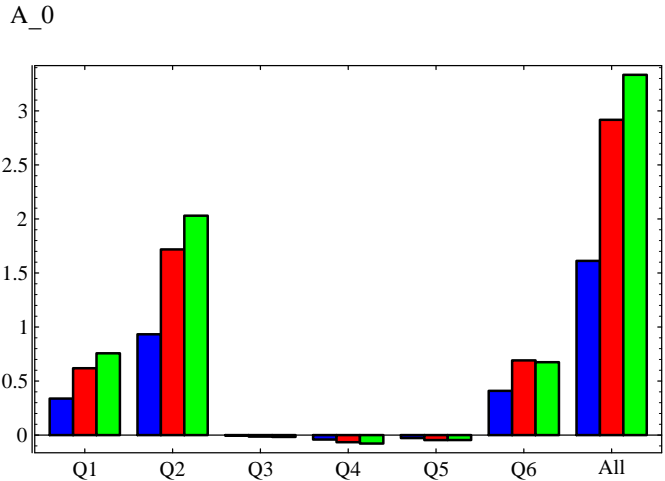}}
\caption{Anatomy of the $A(K^0\to\pi\pi)_{I=0}$ amplitude ($A_0$)
in units of $10^{-7}$ GeV
for central values of the $\chi$QM input parameters\cite{ts98a}:
$O(p^2)$ calculation (black), including minimally subtracted
chiral loops (half-tone), complete $O(p^4)$ result (light gray).}
\label{chart0}
\end{figure}

Fig.~\ref{chart0} shows the individual contributions to the CP conserving
amplitude $A_0$ of the relevant operators, providing us
with a finer anatomy of the NLO chiral corrections.
From Fig.~\ref{chart0} we notice that, because of the chiral loop enhancement,
the $Q_6$ contribution to $A_0$
is about 20\% of the total amplitude. As we shall see,
the $O(p^4)$ enhancement of the $Q_6$ matrix element is what drives \ee 
in the $\chi$QM to the $10^{-3}$ ballpark.

\begin{figure}
\epsfxsize=9cm
\centerline{\epsfbox{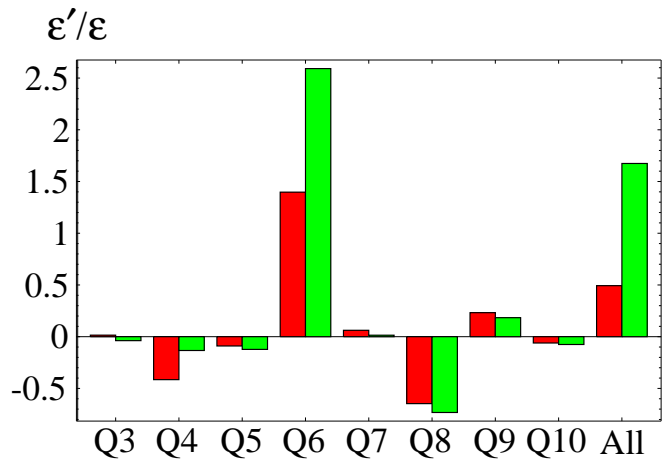}}
\caption{Predicting $\varepsilon'/\varepsilon$:
a (Penguin) Comparative Anatomy of the M\"unchen (dark gray) and 
Trieste (light gray) results (in units of $10^{-3}$).}
\label{istobuqm}
\end{figure}

A commonly used way of comparing the estimates of hadronic matrix elements
in different approaches is via the so-called $B$ factors which
represent the ratio of the model matrix elements to the corresponding VSA
values. However, care must be taken in the comparison of
different models due to the scale
dependence of the $B$'s and the values used by different groups
for the parameters that enter the VSA expressions. 
An alternative pictorial and synthetic
way of analyzing different outcomes for \ee
is shown in Fig.~\ref{istobuqm}, where a ``comparative
anatomy'' of the Trieste and M\"unchen estimates is presented. 

\begin{figure}
\epsfxsize=9cm
\centerline{\epsfbox{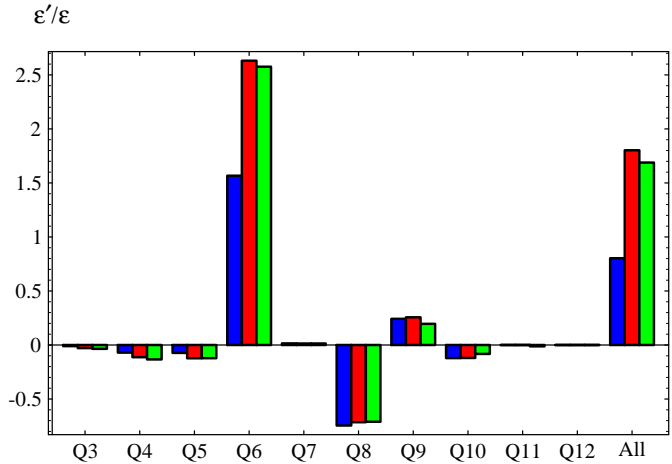}}
\caption{Anatomy of \ee (in units of $10^{-3}$) within the $\chi$QM 
approach\cite{ts98b}. 
In black the LO results (which includes the non-factorizable gluonic
corrections), in half-tone the effect of the inclusion of chiral-loop 
corrections and in light gray the complete $O(p^4)$ estimate.}
\label{charteps}
\end{figure}

From the inspection of the various contributions it is apparent that
the final difference on the central value of \ee is almost entirely
due to the difference in the $Q_6$ component. 
The nature of the $\vev{Q_6}$ enhancement
is apparent in Fig.~\ref{charteps} where the various penguin
contributions to \ee in the Trieste analysis
are further separated in LO (dark histograms) and NLO
components---chiral loops (gray histograms) and tree level
counterterms (dark histograms).

It is clear that chiral-loop dynamics plays a subleading role in the 
electroweak penguin sector ($Q_{8-10}$) while enhancing by 60\% the gluonic
penguin ($I=0$) matrix elements. 

As a consequence, the $\chi$QM analysis shows that the same dynamics 
that is relevant to the
reproduction of the CP conserving $A_0$ amplitude 
(Fig.~\ref{chart0}) is at work
also in the CP violating sector (gluonic penguins).

In order to ascertain
whether these model features represent real QCD effects we must
wait for future improvements in lattice calculations\cite{Martinelli}.
Indications for such a dynamics arise also from $1/N$  
calculations\cite{Hambye} and recent studies of analitic properties
of the $K\to\pi\pi$ amplitudes\cite{dispersive}. 
As a matter of fact, one should expect in general an enhancement of
\ee, with respect to the naive VSA estimate, due to final-state 
interactions. 
In two body decays, the $I=0$ final states feel an 
attractive interaction, of a sign opposite to that of the $I=2$
components. This feature is at the root of the enhancement of the
$I=0$ amplitude over the $I=2$ one. 
 
Recent dispersive analysis\cite{dispersive} of $K\to\pi\pi$ amplitudes
show how a (partial) resummation of final state interactions increases
substantially the size of the $I=0$ components, while slightly depleting
the $I=2$ components.

It is important to notice however that the size of
the effect so derived is generally
not enough to fully account for the $\Delta I = 1/2$ rule.
Other non-factorizable contributions are needed, specially to
reduce the large $I=2$ amplitude obtained from perturbative QCD
and factorization\cite{Cheng}.
In the $\chi$QM approach the fit of the  $\Delta I = 1/2$ rule
is due to the interplay of FSI (at NLO) and non-factorizable soft gluonic
corrections (at LO in the chiral expansion). 

It must be mentioned that
the idea of a connection between the $\Delta I = 1/2$ selection
rule and \ee goes back a long way\cite{VSZacharov&Gilman-Wise},
although at the GeV scale, where we can trust perturbative
QCD, penguins are far from providing
the dominant contribution to the CP conserving amplitudes.

I conclude by summarizing the relevant remarks:

$I=2$ amplitudes:
(semi-)phenomenological approaches which fit the
$\Delta I = 1/2$ selection rule in $K\to\pi\pi$ decays,
generally agree in the pattern and size    
of the $\Delta S = 1$ hadronic matrix elements with the existing
lattice calculations. 

$I=0$ amplitudes:
the $\Delta I = 1/2$ rule forces upon us large deviations from
the naive VSA: $B-$factors of $O(10)$ for $\vev{Q_{1,2}}_0$ 
(lattice calculations presently suffer from large sistematic uncertainties). 

In the $\chi$QM calculation, the fit of the CP conserving
$K\to\pi\pi$ amplitudes feeds down to the penguin sectors
showing a substancial enhancement of the $Q_6$ matrix element, such that
$B_6/B_8^{(2)} \approx 2$. 
Similar indications stem from $1/N$ and dispersive approaches. 
Promising work in progress on the lattice.

Theoretical error:
up to {40\%} of the present uncertainty in the \ee prediction arises
from the uncertainty in the CKM elements {$\Im (V_{ts}^*V_{td})$} 
which is presently controlled by the $\Delta S =2$ parameter {$B_K$}.
A better determination of the unitarity triangle from B-physics
is expected from the B-factories and hadronic colliders\cite{CPprospects}.
From K-physics {$K_L\to\pi^0\nu\bar\nu$} gives the cleanest ``theoretical''
determination of $\Im\lambda_t$\cite{Buchalla}.

{New Physics:} 
in spite of recent clever proposals (mainly SUSY\cite{Masiero})
it is premature to invoke physics beyond the SM
in order to fit \ee.
A number of ungauged systematic uncertainties affect presently all theoretical 
estimates, and, most of all, every attempt to reproduce \ee must also address
the puzzle of the $\Delta I = 1/2$ rule, which is hardly affected by 
short-distance physics. 
Is the ``anomalously'' large \ee 
the ``penguin projection'' of $A_0/A_2 \approx 22$ ?

\end{document}